\documentclass[prb,twocolumn,preprintnumbers,amsmath,amssymb,superscriptaddress,showpacs]{revtex4-1}
\usepackage{graphicx}
\usepackage{amsmath}
\usepackage{amsfonts}
\usepackage{epsfig}
\usepackage{epstopdf}
\usepackage{color}
\usepackage{bm}

\def\beq{\begin{equation}}
\def\eeq{\end{equation}}

\begin{document}

\title{Resonant magneto-optic Kerr effect in the magnetic topological insulator Cr:(Sb$_x$,Bi$_{1-x}$)$_2$Te$_3$}
\author{Shreyas Patankar}
\affiliation{Department of Physics, University of California, Berkeley CA 94720, USA}
\affiliation{Materials Science Division, Lawrence Berkeley National Laboratory, Berkeley CA 94720, USA}
\author{J.\ P.\ Hinton}
\affiliation{Department of Physics, University of California, Berkeley CA 94720, USA}
\affiliation{Materials Science Division, Lawrence Berkeley National Laboratory, Berkeley CA 94720, USA}
\author{Joel Griesmar}
\affiliation{Department of Physics, University of California, Berkeley CA 94720, USA}
\affiliation{Materials Science Division, Lawrence Berkeley National Laboratory, Berkeley CA 94720, USA}
\author{J.\ Orenstein}
\thanks{jworenstein@lbl.gov}
\affiliation{Department of Physics, University of California, Berkeley CA 94720, USA}
\affiliation{Materials Science Division, Lawrence Berkeley National Laboratory, Berkeley CA 94720, USA}
\author{J.\ S.\ Dodge}
\affiliation{Department of Physics, Simon Fraser University, Burnaby, British Columbia, VST 1Z1, Canada}
\author{Xufeng Kou}
\affiliation{Department of Electrical Engineering, University of California, Los Angeles, CA 90095, USA}
\author{Lei Pan}
\affiliation{Department of Electrical Engineering, University of California, Los Angeles, CA 90095, USA}
\author{Kang L.\ Wang}
\affiliation{Department of Electrical Engineering, University of California, Los Angeles, CA 90095, USA}
\author{A.\ J.\ Bestwick}
\affiliation{Department of Physics, Stanford University, Stanford, CA 94305, USA}
\affiliation{Stanford Institute for Materials and Energy Sciences, SLAC National Accelerator Laboratory, 2575 Sand Hill Road, Menlo Park, California 94025, USA}
\author{E.\ J.\ Fox} \affiliation{Department of Physics, Stanford University, Stanford, CA 94305, USA}
\affiliation{Stanford Institute for Materials and Energy Sciences, SLAC National Accelerator Laboratory, 2575 Sand Hill Road, Menlo Park, California 94025, USA}
\author{D.\ Goldhaber-Gordon}
\affiliation{Department of Physics, Stanford University, Stanford, CA 94305, USA}
\affiliation{Stanford Institute for Materials and Energy Sciences, SLAC National Accelerator Laboratory, 2575 Sand Hill Road, Menlo Park, California 94025, USA}
\author{Jing Wang}
\affiliation{Department of Physics, Stanford University, Stanford, CA 94305, USA}
\affiliation{Stanford Institute for Materials and Energy Sciences, SLAC National Accelerator Laboratory, 2575 Sand Hill Road, Menlo Park, California 94025, USA}
\author{Shou-Cheng Zhang}
\affiliation{Department of Physics, Stanford University, Stanford, CA 94305, USA}
\affiliation{Stanford Institute for Materials and Energy Sciences, SLAC National Accelerator Laboratory, 2575 Sand Hill Road, Menlo Park, California 94025, USA}

\begin{abstract}
We report measurements of the polar Kerr effect, proportional to the out-of-plane component of the magnetization, in thin films of the magnetically doped topological insulator $(\text{Cr}_{0.12}\text{Bi}_{0.26}\text{Sb}_{0.62})_2\text{Te}_3$. Measurements of the complex Kerr angle, $\Theta_K$, were performed as a function of photon energy in the range $0.8\text{ eV}<\hbar\omega<3.0\text{ eV}$. We observed a peak in the real part of $\Theta_K(\omega)$ and zero crossing in the imaginary part that we attribute to resonant interaction with a spin-orbit avoided crossing located $\approx$ 1.6 eV above the Fermi energy.  The resonant enhancement allows measurement of the temperature and magnetic field dependence of $\Theta_K$ in the ultrathin film limit, $d\geq2$ quintuple layers. We find a sharp transition to zero remanent magnetization at 6 K for $d<8$~QL, consistent with theories of the dependence of impurity spin interactions on film thickness and their location relative to topological insulator surfaces.
\end{abstract}

\date{\today}


\maketitle

\section{Introduction}

Three dimensional (3D) topological insulators (TIs)~\cite{zhang2011review,hasan2010colloquium} are realizations of quantum matter that manifest protected surface states with Dirac-like energy dispersion. The novel features of TI electrodynamics are expressed in the ``axion" term in the Lagrangian, $L=(\theta/2\pi)(\alpha/2\pi)\mathbf{E}\cdot\mathbf{B}$, where $\theta$ has value the $\pi$ (modulo $2\pi$) in the case of a strong 3D TI~\cite{qi2008topological,essin2009magnetoelectric}, and $\alpha=e^2/\hbar c$ is the fine structure constant. When $\theta$ is uniform in space the axion term has no observable effects because $\mathbf{E}\cdot\mathbf{B}$ is a total space-time derivative and does not modify the equations of motion.  However, observable consequences occur at the interface between topologically trivial and non-trivial media, where a 2D Dirac metal is found.  If the Dirac spectrum is gapped by a time-reversal symmetry (TRS) breaking perturbation, half-integral Hall conductivity, $\sigma_H=\pm e^2/2h$, is expected when the chemical potential lies in the gap~\cite{qi2008topological}.

In practice the half-integral quantum Hall effect (or equivalently, the quantized magnetoelectric effect) is difficult to observe because of parallel current paths that exist either through the bulk or side facets of the crystal that remain gapless despite the breaking of TRS.  As a result, epitaxially-grown thin films are playing an increasing role in exhibiting the electromagnetic phenomena associated with 3D TI~\cite{he2013review}. Particularly interesting physics arises in the transition from 3D to 2D, as realized for example in films in which the thickness can be tuned with atomic precision.  At finite thickness, the top and bottom surfaces are coupled, resulting in two degenerate copies of a massive Dirac spectrum with an energy gap, $\Delta_t$, equal to twice the hopping matrix element, $t$. The development of this gap has been observed directly by \textit{in situ} angle-resolved photo-emission spectroscopy (ARPES)~\cite{zhang2010crossover}. A striking prediction is that, in certain ranges of film thickness, breaking of TRS by ferromagnetic (FM) order will give rise to a quantum anomalous Hall effect (QAHE), that is, integer quantization in the absence of an applied magnetic field~\cite{yu2010quantized}. Several recent reports have confirmed integer quantization with increasing accuracy~\cite{chang2013experimental,kou2014scale,bestwick2014precise}, although under stringent conditions of temperature, chemical doping, and thickness. For example, quantization in TIs doped with the magnetic impurity Cr is observed only for temperatures in the milli-Kelvin range, despite the fact that spontaneous TRS breaking appears at 15-20~K.

To better understand the fragility of the QAHE in magnetic-impurity doped epitaxial films, it is important to probe the strength of the FM order as a function of film thickness $d$, temperature $T$, and magnetic field $B$. In this work we measure the polarization rotation of light on reflection at normal incidence (Kerr effect), which vanishes in the presence of TRS~\cite{halperin1992local} and is proportional to the out-of-plane component of the magnetization, $M_z$, when TRS is broken. We report measurements of the real and imaginary parts of the complex Kerr angle, $\Theta_K$, from near-IR through visible wavelengths (photon energies in the range 0.8 eV$<\hbar\omega<$ 3.0 eV). Although, in principle, detecting $\Theta_K(\omega)$ on the energy scale of gap induced by FM order, $\Delta_M$, might be preferable in terms of direct contact with theoretical predictions~\cite{lasia2014optical,tse2011magneto,tse2010giant,wang2013magneto}, in practice the present limits on sensitivity in this spectral range make this approach difficult. The magnitude of the Kerr angle predicted theoretically is $\sim\alpha/(\epsilon-1)\sim 1$ mrad, where $\epsilon$ is the bulk (or substrate in the case of an ultrathin film) dielectric constant. Thus, for $\Theta_K$ to be an effective probe of $\Delta_M$, particularly the manner in which it vanishes in the limit that $d,T,B\rightarrow 0$, requires $\sim$10-100 $\mu$rad sensitivity in the frequency regime of 2-20 THz. This can be compared with the current state of the art of terahertz spectroscopy, where sensitivity $\sim$0.1 mrad in a frequency range up to $\sim$ 2 THz has been demonstrated in measurements on Bi$_2$Se$_3$ films~\cite{aguilar2012terahertz}.

In the the near-infrared through visible range of our spectrometer, we observe a peak in the real part of $\Theta_K(\omega)$ and a zero crossing in the imaginary part at approximately 1.7 eV that we attribute to optical transitions involving states near a spin-orbit avoided crossing in the bandstructure.  The enhanced Kerr rotation on resonance allows measurement of $\Theta_K$ in the ultrathin film limit, $d\geq2$ quintuple layers. We find that the FM order decreases rapidly with decreasing thickness, which we attribute to the combined effects of disorder inherent in a doped magnetic semiconductor as well as competing anisotropic exchange interactions resulting from strong spin-orbit coupling. Finally, additional evidence for disordered magnetism is found in the significant difference in $\Theta_K(T)$ observed on warming in zero as compared with 100 mT applied magnetic field.

\section{Optical methods, material synthesis and characterization}

The samples used for this study are thin films of $(\text{Cr}_{0.12}\text{Bi}_{0.26}\text{Sb}_{0.62})_2\text{Te}_3$ prepared using molecular-beam epitaxy on semi-insulating GaAs (111) substrates. These samples are not gated, so their Fermi level is determined by film stoichiometry. This is not perfectly controlled during growth, and can further change with exposure to air, yielding a range of carrier concentrations for the different samples from $-8\times10^{12}$ (i.e., hole doped) to $9\times10^{12}$ cm$^{-2}$ (electron doped), as determined from Hall measurements at temperatures above the magnetic transition~\cite{SuppInfo}. In a subset of the samples studied the anomalous Hall resistance $R_{\mathrm{AHE}}$ approaches the quantized value $h/e^2\approx 25.8~$k$\Omega$ at dilution refrigerator temperatures~\cite{kou2014scale,bestwick2014precise}, while other samples grown under nominally the same conditions display $R_{\mathrm{AHE}}$ in the range $2$ to $17$~k$\Omega$. Despite these variations in $R_{\mathrm{AHE}}$, we find that the trends in $\Theta_K(d,T,B)$ are reproducible. This suggests that the Kerr effect in the near-IR, in contrast to low temperature transport, does not originate principally from surface states. In this case reproducibility of $\Theta_K$ requires only that the chemical potential remain in the bulk gap, which is a much less stringent condition than required for observation of the QAHE.

The Kerr spectra reported here were measured with a W-filament light source and grating monochromator, and single wavelength temperature and magnetic field scans were performed with a HeNe laser source at 633~nm (1.96~eV). In a polar Kerr measurement, linearly polarized light is normally incident on a sample and the reflected light emerges in an elliptical polarization state whose major axis is rotated with respect to the initial polarization direction. The real and imaginary parts of $\Theta_K$ correspond to rotation and induced ellipticity, respectively. To measure the full complex $\Theta_K$ we used a technique based on a photo-elastic optical phase modulator (PEM)~\cite{badoz1977sensitive,sato1981measurement}. Synchronous detection of the reflected light at the second and third harmonic of the PEM frequency enabled precise discrimination between the real and imaginary parts of $\Theta_K$, respectively, as well as detection with 10 $\mu$rad sensitivity.
\begin{figure}[htbp]
\begin{center}
\includegraphics[width=3.3in]{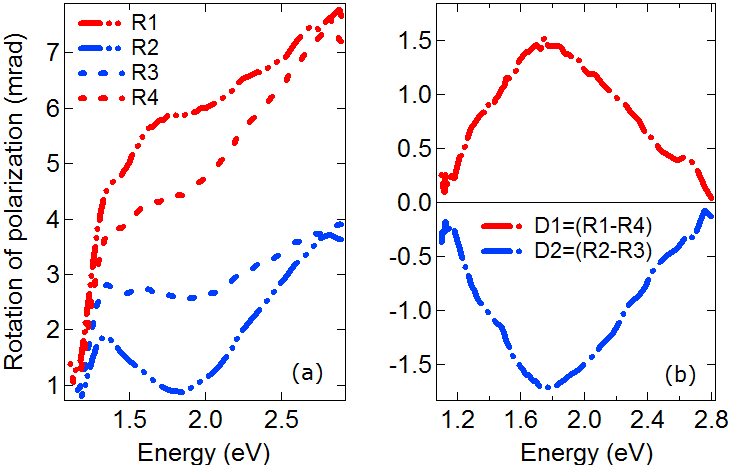}
\end{center}
\caption{Isolating the magneto-optic Kerr effect at $6$~K for $20$~QL sample. (a) $\Theta_K$ at at 6 K in $+100$~mT (R1), and $-100$~mT (R2), and at $50$~K in field $-100$~mT (R3) and $+100$~mT (R4). (b) $\Theta_K$ at $50$~K subtracted from $\Theta_K$ at $6$~K in field $+100$~mT (D1) and $-100$~mT (D2).}
\label{fig1}
\end{figure}

By itself, this method does not distinguish between rotation induced by TRS breaking and strain birefingence of the sample and cryostat window. In addition, in the presence of an applied magnetic field there is a contribution to the measured rotation from the Faraday effect of the window. To isolate the magneto-optic Kerr effect of the sample we use the following procedure, which is illustrated in Fig.~\ref{fig1}. We first measure the spectrum of $\Theta_K$, in both positive and negative magnetic fields, at a temperature such as 50~K where the magnetic response is well below our detection limit. We then measure $\Theta_K$ with the sample cooled to the desired temperature, for example $6$~K, again for both positive and negative magnetic fields and calculate the difference spectra, $\Theta_K(6\mathrm{K}, +B)-\Theta_K(50\mathrm{K}, +B)$ and $\Theta_K(6\mathrm{K}, -B)-\Theta_K(50\mathrm{K}, -B)$ (shown in Fig.~\ref{fig1}b). Since the sample is cooled independently of its environment, this step eliminates contributions from Faraday rotation and strain birefringence of the cryostat window (which remains at room temperature). Finally, we calculate the component of the spectra in Fig.~\ref{fig1}b that is odd in applied magnetic field to obtain the  magneto-optic Kerr spectra shown in succeeding figures of this paper.

\section{Results}

\subsection{Complex Kerr angle and conversion to off-diagonal dielectric function}

\begin{figure}[t]
  \centering
    \includegraphics[width=3.5in]{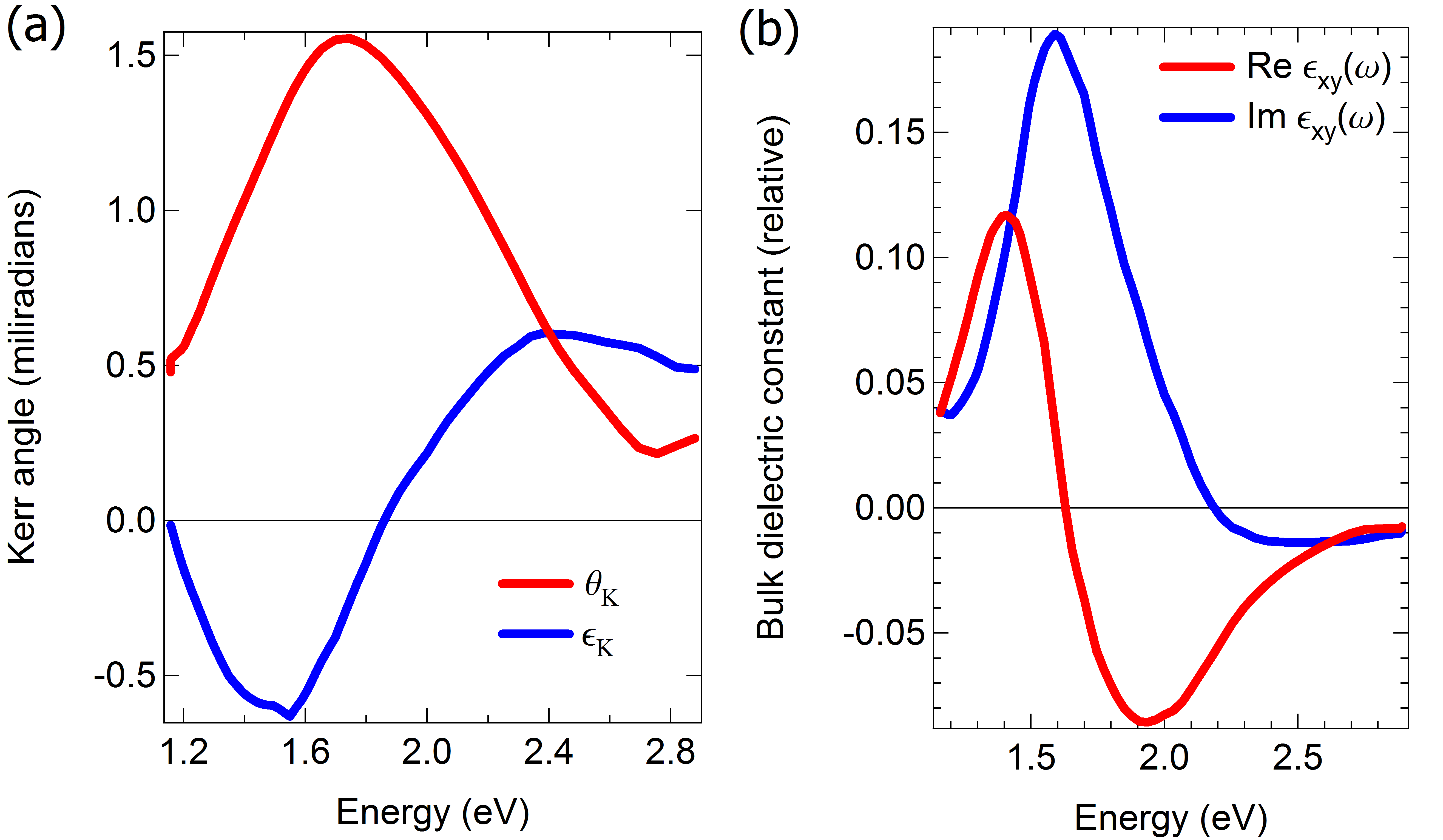}
  \caption{Kerr effect in a 20~QL thick sample of (Cr$_{0.12}$Bi$_{0.26}$Sb$_{0.62}$)$_2$Te$_3$. (a) Real $\theta_K$ and imaginary $\varepsilon_K$ parts of the Kerr spectrum measured at 6~K with magnetic field of 100~mT applied perpendicular to the film plane, (b) off-diagonal component of the dielectric function, $\epsilon_{xy}(\omega)$, as determined by analysis of the Kerr spectra shown in panel (a).}
  \label{fig2}
\end{figure}

Fig.~\ref{fig2}a shows the real, $\theta_K$, and imaginary, $\varepsilon_K$ (not to be confused with the dielectric constant $\epsilon$), parts of $\Theta_K$ in the spectral range from 1.1-2.9~eV for a 20 quintuple-layer (QL) sample measured at $T=6$~K using the procedure described above. We find a resonance in the Kerr spectrum, characterized by a peak in the real part and a zero crossing in the imaginary part. In the following we describe the extraction of the off-diagonal component of the complex dielectric tensor, $\epsilon_{xy}(\omega)$ from the measured $\Theta_K(\omega)$.

Because the optical penetration depth is less than 10~QL over most of the spectral range of our measurement, we model the 20~QL sample as a semi-infinite medium and assume that $\epsilon_{ij}(\omega)$ is uniform throughout the film. In this case,
\beq
\epsilon_{xy}(\omega)=\frac{\tilde{n}(\tilde{n}^2-1)}{2}\Theta_K(\omega),
\eeq
where $\tilde{n}^2(\omega)\equiv\epsilon_{xx}(\omega)$. In the analysis we assume that $\epsilon_{xy}\ll\epsilon_{xx}$ and use the diagonal component of the dielectric function as reported in Ref.~\onlinecite{park2008optical}.

Fig.~\ref{fig2}b shows the real and imaginary parts of $\epsilon_{xy}(\omega)$ obtained by analysis of $\Theta_K(\omega)$. The appearance of a peak in $\epsilon_{1xy}(\omega)\equiv\mathrm{Re}[\epsilon_{xy}(\omega)]$ and dispersive lineshape of $\epsilon_{2xy}(\omega)\equiv\mathrm{Im}[\epsilon_{xy}(\omega)]$ are clear evidence for a resonance at approximately 1.6~eV.  Note that the appearance of a peak in the imaginary part and dispersive lineshape in the real part of $\epsilon_{xy}(\omega)$ is reversed compared with the usual structure of $\epsilon_{xx}(\omega)$ in the neighborhood of a resonance.  This reversal is seen as well in spectra obtained from the magnetically doped 3D semiconductor Mn:GaAs~\cite{kojima2003observation,sun2011above} and is consistent with what is expected for Kerr spectra in dilute magnetic semiconductors.  As a consequence of TRS breaking the energy of transitions between valence and conduction bands is slightly different for left and right circularly polarized light.  Thus the difference spectrum, $\epsilon_{2+}(\omega)-\epsilon_{2-}(\omega)$ (where $\epsilon_{2\pm}$ are the imaginary parts of the dielectric function in the circular polarized basis) will resemble the derivative of an absorption peak. This quantity, the \textit{imaginary} part of $\epsilon_+(\omega)-\epsilon_-(\omega)$, is directly proportional to the \textit{real} part of $\epsilon_{xy}(\omega)$ because $\epsilon_{xy}(\omega)=[\epsilon_{+}(\omega)-\epsilon_{-}(\omega)]/2i$ on transforming back to the Cartesian basis. Thus the derivative-like lineshape of $\epsilon_{2xy}(\omega)$ that we observe is consistent with expectations for interband transitions in 3D dilute magnetic semiconductors.

\begin{figure}[t]
  \centering
    \includegraphics[width=1.7in]{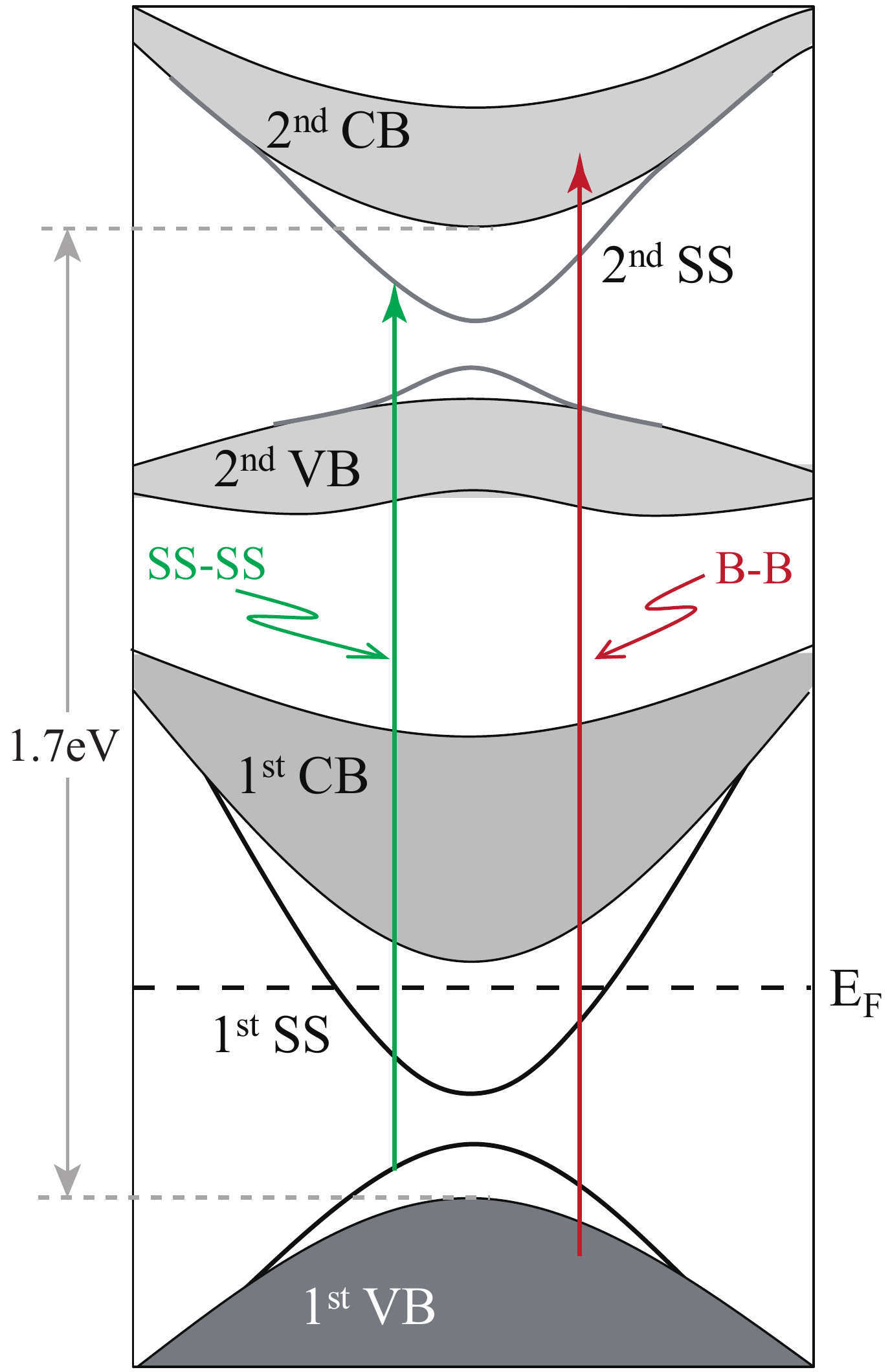}
  \caption{Schematic illustration of the band structure and optical transitions in magnetic TI $(\text{Cr}_{0.12}\text{Bi}_{0.26}\text{Sb}_{0.62})_2\text{Te}_3$. SS, VB, and CB denote surface state surface,  valence band, and conduction band, respectively. The shaded region in CB and VB indicate the energy range of two-dimensional subbands. The SS gap is induced by FM order. The green line illustrates the first SS to second SS optical transition, while the red line illustrates transitions between subband states that evolve to the  ``bulk-to-bulk'' (BB) transition with increasing film thickness.}
  \label{fig3}
\end{figure}

Two-photon ARPES measurements~\cite{sobota2013direct} as well as bandstructure calculations~\cite{niesner2012unoccupied,eremeev2013new}, suggest that the resonance at 1.6~eV is associated with transitions from states near the Fermi energy, $\varepsilon_F$, to states near a spin-orbit avoided band-crossing situated at the $\Gamma$ point at approximately $1.5$-$1.8$~eV above $\varepsilon_F$. Transitions resonant with spin-orbit avoided band crossings are known to make a large contribution to $\epsilon_{xy}(\omega)$~\cite{fang2005orbital}. A schematic bandstructure diagram for a film of thickness 20~nm, showing the avoided crossings at the $\Gamma$ point is shown in Fig.~3. The upper avoided crossing is a higher energy replica of the band inversion that takes place near the Fermi energy, and is associated as well with a symmetry predicted Dirac surface-state band. The shaded regions in the schematic illustrate the range of energies associated with confinement in the direction normal to the plane of the film and the vertical arrow shows a typical transition from subband states of the first valence band (VB1) to the second conduction band (CB2).

\subsection{Film thickness dependence}

\begin{figure}[t]
  \centering
    \includegraphics[width=3.55in]{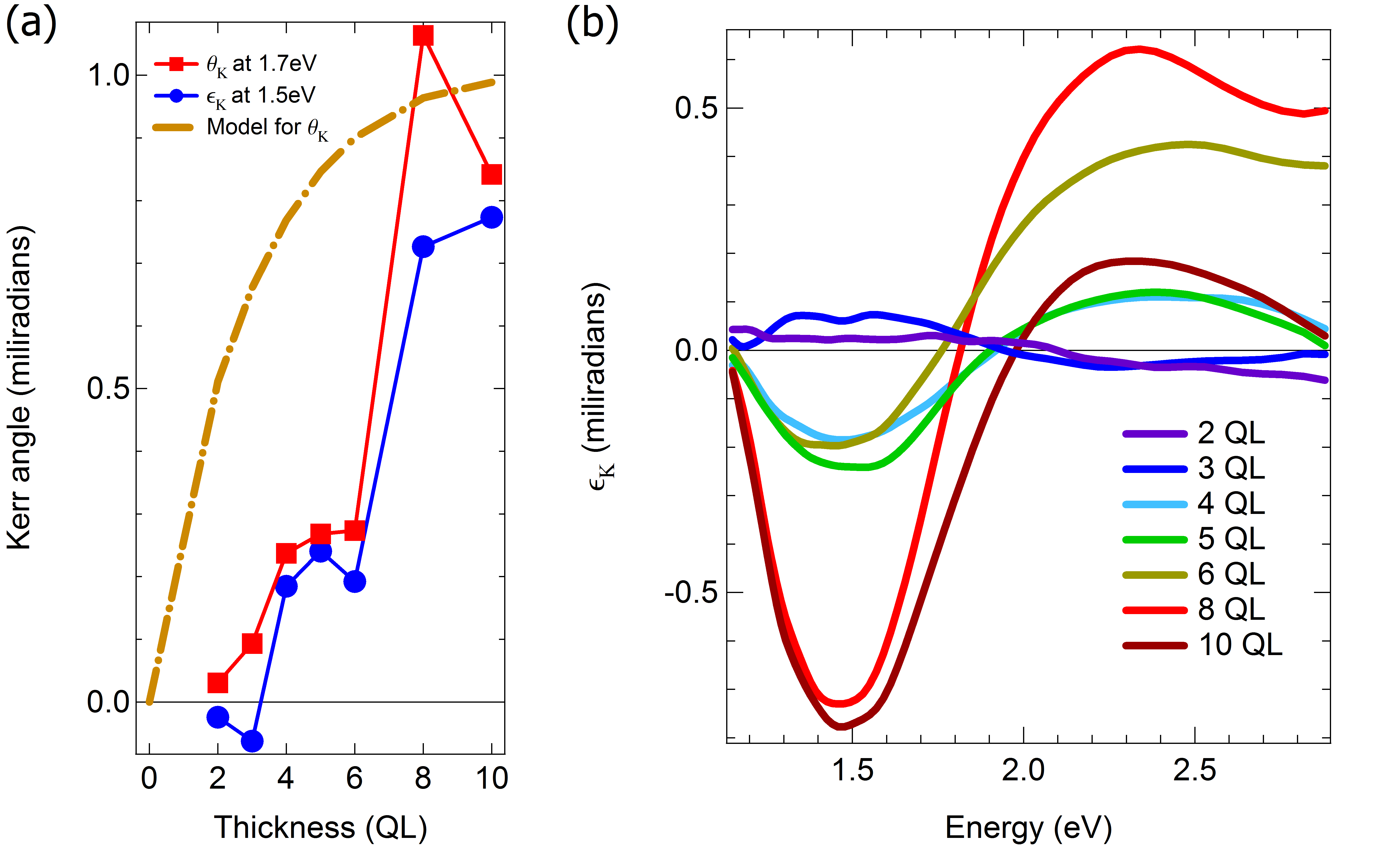}
  \caption{Thickness dependence of Kerr angle. (a) Amplitudes of $\theta_K$ and $\varepsilon_K$ at their respective peaks of 1.7~eV and 1.5~eV as a function of sample thickness at 6~K and 100~mT. Plotted for comparison (dashed-dot line) is $\theta_K(d)$ calculated under the assumption of $d$-independent $\epsilon_{xy}(\omega)$, (b) imaginary part of $\Theta_K(\omega)$ for various thicknesses in the range from 2 to 10~QL.}
  \label{fig4}
\end{figure}

We use the sensitivity of the resonantly enhanced Kerr effect to measure the Kerr angle as a function of film thickness, $d$, down to the ultrathin limit. Fig.~\ref{fig3}a shows the real and imaginary parts of $\Theta_K$ at $T=6$~K as a function of $d$, with fixed Cr concentration, sampled at their peak photon energies of 1.7 and 1.5~eV, respectively. Fig.~\ref{fig4}b shows the imaginary part of $\Theta_K(\omega)$ for the same set of samples. We observe that the frequency dependence of the Kerr resonance remains unchanged even as its amplitude is significantly reduced with decreasing $d$. The insensitivity of the spectrum to thickness suggests that the width of the resonance is largely determined by the dispersion of energies with respect to the in-plane wavevector, rather than the energies associated with subband splittings.  This conclusion is consistent with bandstructure calculations that indicate subband splitting are no larger than 0.2~eV for films as thin as 4~nm~\cite{liu2010oscillatory,lasia2014optical}, which can be compared with the characteristic width of the resonance in $\Theta_K(\omega)$, which is approximately 0.8~eV.

The dependence of the amplitude of the Kerr angle on thickness is highly structured, with steep changes in amplitude taking place just below 8 and 4~QL and relatively small changes in between these thicknesses.  These features can be contrasted with the thickness dependence that would obtain if $\epsilon_{xy}(\omega)$ was independent of sample thickness. To make this comparison, we fitted $\epsilon_{xy}(\omega)$ as measured from the 20~QL film using a Lorentzian function, and then input this fit into the formula for the reflection tensor of a thin film on a dielectric substrate~\cite{tse2011magneto}. The result of this simulation is shown as a dashed-dot line in 
Fig.~\ref{fig3}a.  Not surprisingly, the assumption of thickness independent $\epsilon_{xy}(\omega)$ predicts that the Kerr rotation decreases smoothly to zero once $d$ becomes smaller than the optical penetration depth. Thus the sharp features in the observed thickness dependence can be attributed to structure in $\epsilon_{xy}(\omega)$ vs. $d$, and ultimately to structure in $M_z(d)$.

\subsection{Temperature dependence}

\begin{figure}[b]
  \centering
    \includegraphics[width=3.3in]{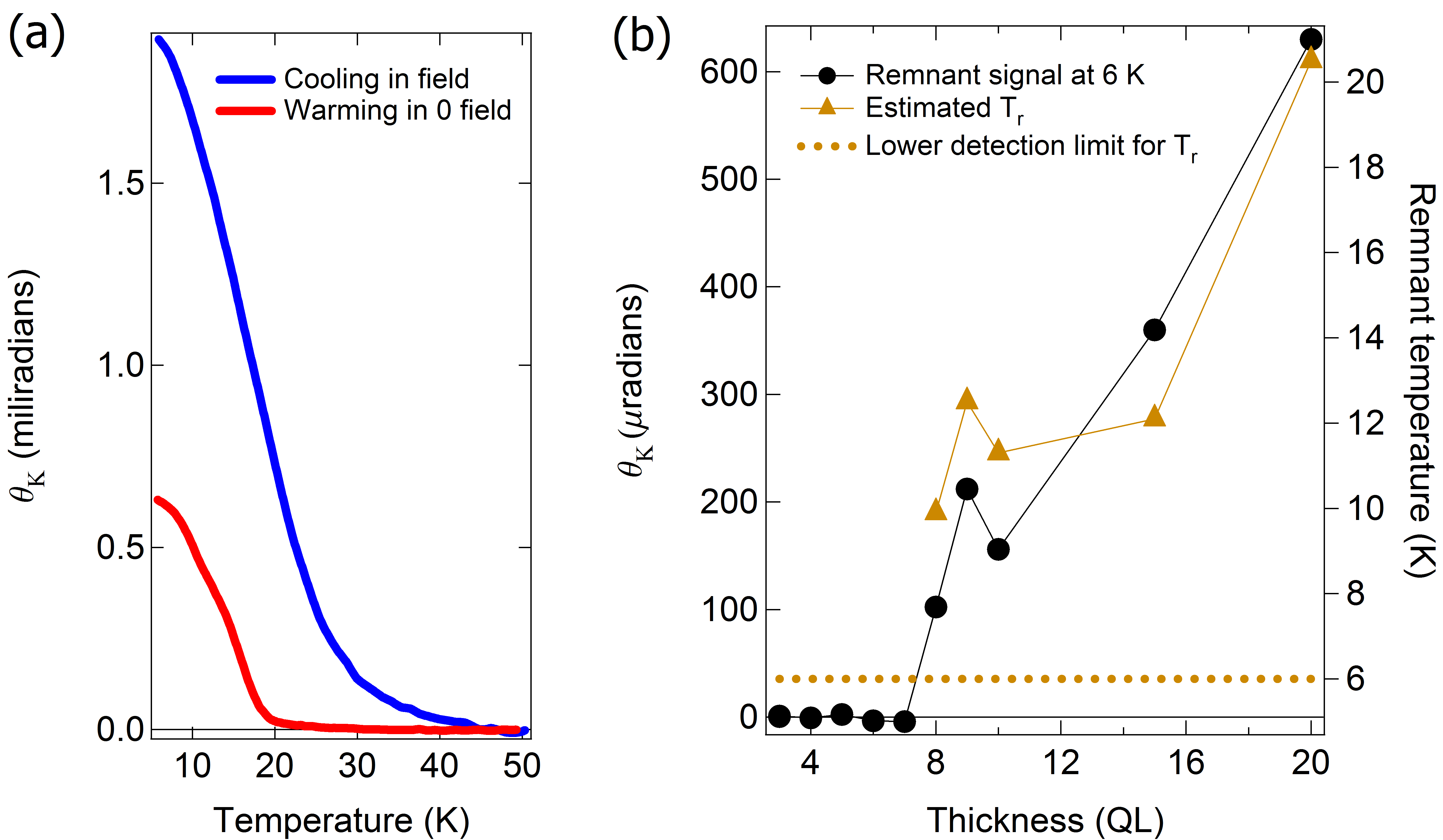}
  \caption{(a) Real part of the Kerr rotation, $\theta_K$ at photon energy 1.96~eV as a function of temperature for a 20~QL film. The larger amplitude curve was measured upon cooling in a 100~mT magnetic field and warming with the field left on, the smaller amplitude curve was recorded on warming with field turned off at 6~K, and is a measure of the remnant magnetization, (b) Remanent $\Theta_K$ at 6~K as a function of film thickness shown as solid circles(left axis); triangles (right axis) show the temperature T$_r$ above which the remanent signal vanishes. Dashed line indicates our experimental upper bound on $T_r$.}
  \label{fig5}
\end{figure}

Further evidence for structure in $M_z(d)$ emerges from the $T$ dependence of the Kerr rotation. Fig.~\ref{fig5}a shows $\theta_K(T)$ of a 20~QL film at photon energy 1.96~eV measured with two different protocols for application of the $B$ field. In both measurements the sample was cooled to 6~K with $B\approx100$~mT applied in the direction normal to the film surface. The data shown in red were recorded after switching the field to zero at 6K and subsequently warming and are therefore proportional to the remanant magnetization. We define the temperature at which $\theta_K(T)$ warmed in zero field extrapolates to zero as the remnant temperature, $T_r$, which in this case is $\sim$~20K. The data shown in blue were recorded after again cooling in $100$~mT, but with the field left on during the warming cycle. The different results for the two measurement procedures are quite striking and not characteristic of a conventional ferromagnet. Not only is $\theta_K(T)$ in field much larger, but it persists to much higher temperature, $\sim$~50K. Thus there is a broad temperature regime, from $\approx20-50$~K, in which the magnetization in zero-field is absent but a magnetization comparable to the remanent value is induced by magnetic fields in the 100~mT range.

In Fig.~\ref{fig5}b we illustrate the thickness dependence of the remanent magnetization by plotting both $T_r$ and the remanent value of $\Theta_K(6\mathrm{K})$ \textit{vs}. $d$. The structured dependence on $d$ is even more clear in the remanent properties than those measured in 100~mT magnetic field. For example, $\Theta_K(6\mathrm{K})$ suddenly becomes too small to measure for sample thicknesses less than 8~QL. We note that samples such as these may exhibit remnant magnetization when cooled below 6 K, for example into the miliKelvin range where the QAHE has been observed. Below we comment on possible origins for such structure in $M_z(d)$ and $T_r(d)$.

\section{Discussion}

Evidence for a crossover to a regime of strong coupling between top and bottom surface states with decreasing film thickness was reported in a time-domain terahertz study~\cite{wu2013sudden} of the system (Bi$_{(1-x)}$In$_x)_2$Se$_3$. Analysis of the real and imaginary parts of the Drude conductivity indicated a crossover thickness, $d_c(x)$, below which the carrier scattering rate, $1/\tau$ increased rapidly. The crossover was associated with entering a regime where $\Delta_t(d)>1/\tau$, which is the condition for loss of topological protection against backscattering. For $x=0$, $d_c$ is $\sim$ 6 QL, which is comparable to, but somewhat smaller than, the thicknesses at which the remanent $\Theta_K$ changes abruptly in our somewhat similar but magnetically-doped films. That the crossover to strong coupling between top and bottom surfaces, as heralded by $\Theta_K$ dropping to zero, occurs at a larger thickness in the magnetic film is consistent with the expected reduction spin-orbit coupling strength that results from Cr-doping~\cite{zhang2013,kou2015mapping}. The smaller spin-orbit coupling reduces the inverted bulk band gap, consequently increasing the penetration depth into the bulk of Dirac surface states. In the earlier study, below $d_c$ the terahertz conductance decreases smoothly but extrapolates to zero at $d\approx 2 $~QL. The absence of conductance at 2~QL was associated with either the theoretically expected transition to a topologically trivial state, or strong disorder in the ultrathin limit. As shown in Fig.~2, $\Theta_K(\omega)$ shows an analogous transition to zero response at this same thickness.

The pronounced structure that we observe in $\Theta_K(d)$ likely reflects a strong dependence of spin interactions on film thickness. This dependence, in turn, is a consequence of the complicated spin interactions that exist even in the thick film regime. Theoretical studies~\cite{liu2009magnetic,zhu2011electrically,abanin2011ordering} have found that the effective interaction between impurity spins, as mediated by TI surface state electrons, is highly anisotropic because of the spin-momentum locking of the DBs. The interaction is characterized by an Ising term, $U_{12}=-J_zS_1^zS_2^z$, that favors FM alignment in the direction normal to the film plane as is experimentally observed in our thicker films, plus contributions that depend on the relative position, $\bf{r}$, of the spins.  One such contribution favors FM alignment in the direction parallel to $\bf{r}$ and the other favors anti-FM (AFM) alignment in the direction perpendicular to $\bf{r}$.  If the position-dependent terms are sufficiently weak, the ground state has FM order along $z$.  However, if the position-dependent interactions are strong enough the ground state will be a spin glass~\cite{abanin2011ordering}.

Two recent theoretical studies have considered the dependence of these competing interactions on film thickness~\cite{litvinov2014magnetic,wang2014}. As pointed out in Ref.~\onlinecite{wang2014}, the Ising FM interaction arises mainly from excitations across the gap, and therefore decreases monotonically with $d$ due to increased $\Delta_t(d)$. On the other hand, the AFM interactions, which derive from superexchange, depend only weakly on the gap, and consequently on thickness.  Therefore, as films become sufficiently thin, the AFM interaction will eventually dominate, in which case the random distribution of magnetic impurities~\cite{lee2015} will frustrate an AFM ground state and lead to paramagnetic or spin glass behavior.

A similar conclusion was reached in Ref.~\onlinecite{litvinov2014magnetic}, in which the dependence of the effective interaction on the distance of the impurities from the surface was included.  In that work it was reported that, while the Ising interaction mediated by a single surface state is FM, in agreement with previous work~\cite{zhu2011electrically,liu2009magnetic,abanin2011ordering}, the interaction between spins situated near opposite surfaces is AFM. For a film that is sufficiently thin, interactions that favor alignment along $z$ can be frustrated by positional disorder and the net weakening of the Ising contribution to the exchange energy can again lead to spin glass behavior. Our observation that as opposite surfaces approach each other the remanent magnetization vanishes, while the film retains a large susceptibility to external fields, is consistent with both the above theoretical pictures.

In summary, we have measured the complex magneto-optic Kerr effect in $(\text{Cr}_{0.12}\text{Bi}_{0.26}\text{Sb}_{0.62})_2\text{Te}_3$ in visible and near-infrared wavelengths as a function of temperature, photon energy and magnetic field. We observed a resonance in the Kerr effect at 1.7~eV, corresponding to the energy of a spin-orbit avoided band crossing with respect to the Fermi energy. The resonant enhancement allowed us to explore the effective out-of-plane component of the magnetization in the ultrathin film limit. At our base temperature of 6~K, we observed a transition to zero remanent magnetization below a thickness of 8~QL, and to essentially zero magnetism at 2~QL. The dependence of $M_z$ on $d$ is consistent with theories of thickness dependence of magnetic impurity spin interaction in TIs.

\begin{acknowledgments}
Research primarily supported by the U.S. Department of Energy, Office of Science, Basic Energy Sciences, under Contract No.~DE-AC02-76SF00515. J. G. acknowledges a scholarship from the Materials Science and Active Surface Program at Ecole Polytechnique, Palaiseau, France (Chaire X-ESPCI-Saint-Gobain) for support. K.~L.~W. acknowledges the support of the Raytheon endorsement. A.~J.~B. acknowledges support from a Benchmark Stanford Graduate Fellowship. E.~J.~F. acknowledges support from a DOE Office of Science Graduate Fellowship. Materials growth, surface characterization, preliminary electrical characterization, and electronic instrumentation were supported by the DARPA MESO program under Contracts No.~N66001-12-1-4034 and No.~N66001-11-1-4105. Infrastructure and cryostat were funded in part by the Gordon and Betty Moore Foundation through Grant GBMF3429 to D. G.-G.
\end{acknowledgments}

\end{document}